\documentclass[aps,prb,twocolumn,amssymb,amsfonts,groupedaddress,showpacs]{revtex4}
\usepackage{epsfig}
\usepackage{amsmath}
\usepackage{bm}
\usepackage{times}

\begin{document}

\title{\normalsize\textbf{{Dynamical nuclear spin polarization and the Zamboni
effect in gated double quantum dots}}}

\author{Guy Ramon}
\email{guy.ramon@gmail.com}
\author{Xuedong Hu}
\affiliation{Department of Physics, University at Buffalo, SUNY, Buffalo, NY
14260-1500}

\begin{abstract}
A dynamical nuclear polarization scheme is studied in gated double
dots. We demonstrate that a small polarization ($\sim 0.5\%$) is
sufficient to enhance the singlet decay time by two orders of
magnitude.
%
%
This enhancement is attributed to an equilibration process between
the nuclear reservoirs in the two dots accompanied by reduced
fluctuations in the Overhauser fields, that are mediated by the
electron-nuclear spin hyperfine interaction.
\end{abstract}


\maketitle

Electron spins localized in semiconductor quantum dots have been
intensively investigated in recent years due to their potential use
in quantum information processing and spintronics.\cite{ZFD}
%
%
Several experimental \cite{Johnson,Koppens,Petta} and theoretical
\cite{Merkulov,Schliemann,Shenvi,Coish,Deng,Yao} studies have
identified the hyperfine (HF) interaction between an electron spin
and the surrounding nuclear spins as one of the main sources for
electron spin decoherence in low temperature GaAs quantum dots,
leading to $T_2^*$ on the order of $10-25$ns and $T_2$ on the order
of $\mu$s. These values are orders of magnitude shorter than the
spin relaxation time, which approaches tens of milliseconds in these
systems.\cite{Johnson}
%

Several strategies have been suggested to alleviate electron spin
decoherence via nuclear spins, including spin-echo techniques (to
remove inhomogeneous broadening \cite{Petta,Shenvi,Yao}), nuclear
spin state measurement (to narrow the Overhauser field
distribution,\cite{Klauser}) and nuclear polarization (to reduce
phase space for nuclear spin dynamics.\cite{Burkard})  Nuclear spin
polarization is also valuable for state initialization in NMR
quantum computing and for utilizing collective nuclear states as
long-lived quantum memory.\cite{Taylor1} So far, optical pumping has
produced up to $\sim 60\%$ nuclear polarization \cite{Gammon} in
interface fluctuation GaAs dots, while spin transfer via hyperfine
mediated spin-flip scattering in the spin-blockade regime in gated
GaAs dots has led to $\sim 1\%$ polarization.\cite{Ono,Marcus} Among
the limiting factors in these dynamical polarization schemes are the
large difference in the Zeeman energies of the electrons and nuclei
making joint spin flip processes energetically unfavorable, and
nuclear spin diffusion due to dipolar interaction.  In contrast,
theoretical studies have shown that in order to achieve a sizeable
enhancement of single electron spin decoherence time via phase-space
squeezing, a nuclear polarization of more than 99\% is
required.\cite{Sousa,Coish,Deng}

In this paper we show that HF interaction can be exploited to
dynamically polarize the nuclear spins in gated double dots.  Most
interestingly the relaxation time of the two-electron spin singlet
state dramatically increases without the need for nearly complete
nuclear polarization.  This suppression of relaxation is achieved by
an equilibration process in the nuclear reservoirs in the two dots
and a reduction in the fluctuations of their Overhauser
fields.  We have dubbed this effect, mediated by the HF interaction
during the polarization cycles, as the nuclear Zamboni
effect.\cite{Zamboni}
%

\begin{figure}[htbp]
\epsfxsize=0.8\columnwidth
\vspace*{0.2cm} \centerline{\epsffile{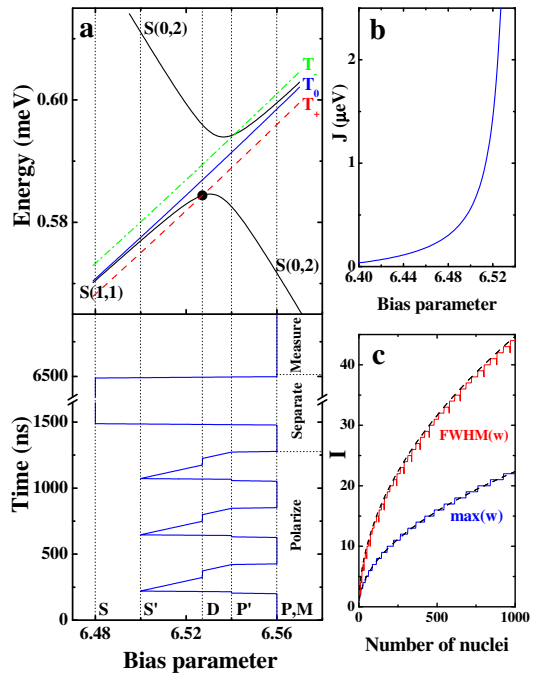}} \vspace{-0.2cm}
\caption{(Color online) (a) Upper panel: Orbital energy diagram for
the double dot near the (1,1)-(0,2) transition vs. a bias parameter,
proportional to the inter-dot bias gate potential. Shown are the
hybridized singlet states (black curves), and split (1,1) triplet
states $T_-$ (dash-dotted green), $T_0$ (blue), and $T_+$ (dashed
red). $\bullet$ denotes the $S-T_+$ degeneracy point.
Lower panel: Polarize, separate and measure pulse sequence. Three
polarization cycles are shown followed by a separation time and a
measurement of the singlet probability. The letters at the bottom
indicate biasing points discussed in the main text. (b) Exchange
energy as a function of bias. (c) Center location (blue) and FWHM
(red) of the nuclear spin state distribution as a function of $N$.
The dashed lines are $\sqrt{N/2}$ and $\sqrt{2N}$, corresponding to
the center location and FWHM of the distribution, respectively.}
\label{Fig1}
\end{figure}

We study the dynamics of a system of two electrons localized in a
gated double dot interacting with two nuclear spin baths within the
framework of the Hamiltonian
\begin{equation}
H=H_{\rm orb}+H_Z +H_{\rm HF},
\label{Ham}
\end{equation}
where we have neglected nuclear-nuclear dipolar coupling in the
current study. For the orbital part we adapt the Hund-Mulliken
approach \cite{Burkard,HD} to solve for the electronic states in the
gated dot configuration, where $H_{\rm orb}$ includes the
single-particle Hamiltonian and the Coulomb interaction. The
relevant Hilbert space of $H_{\rm orb}$ is spanned by four
two-particle states, $\{S(2,0),S(0,2),S(1,1),T(1,1)\}$, which
consist of the separated singlet and triplet and the two doubly
occupied singlet states. Indices $(i,j)$ indicate the number of
confined electrons in the (left,right) dot. We neglect the doubly
occupied triplet states as their energy is much higher for the
structures under study.\cite{Johnson,Petta} The Zeeman interaction
$H_Z=g \mu_B {\bf B} \cdot \sum_{i=L,R} {\bf S}_i$, with $g=-0.44$
and $\mu_B$ being the Bohr magneton, splits the triplet states.  The
resulting energy diagram and exchange interaction $J$ near the (1,1)
to (0,2) charge transition are shown in Figs.~\ref{Fig1}a,b for
$B=100$ mT ($E_Z=2.5 \mu {\rm eV}$), half interdot distance
$a=1.9a_B$, and dot confinement $\omega_0=120 \mu {\rm eV}$,
corresponding to the experimental parameters in
ref.~\onlinecite{Petta}.

For the HF interaction we consider the contact term between the two
electrons and the surrounding nuclei
\begin{equation}
H_{\rm HF}=\sum_{i=L,R} \sum_k A^k_i {\bf I}^k_i \cdot {\bf S}_i
={\bf h} \cdot {\bf S}+\mbox{\boldmath $ \delta {\rm h}$} \cdot
\mbox{\boldmath $ \delta {\rm S}$} \,,
\label{HFham}
\end{equation}
where $A_i^k$ is the HF coupling constant with the $k$th nucleus in
the $i$th dot.  Here ${\bf h}=\frac{1}{2}\left({\bf h}_L+{\bf
h}_R\right)$, $\mbox{\boldmath $ \delta {\rm
h}$}=\frac{1}{2}\left({\bf h}_L-{\bf h}_R\right)$ and ${\bf S}={\bf
S}_L+{\bf S}_R$, $\mbox{\boldmath $ \delta {\rm S}$}={\bf S}_L-{\bf
S}_R$ are the sums and differences of the nuclear fields and the
electron spins in the two dots, respectively, where ${\bf h}_i =
\sum_k A_i^k {\bf I}^k_i$ is the nuclear field in dot $i$.  We
assume $I_i^k=\frac{1}{2}$ for simplicity.  The Hamiltonian in
Eq.~(\ref{Ham}) conserves the total spin and can be block
diagonalized in each of the Hilbert subspaces defined by the
eigenvalues of the operator $J_z=S_L^z+S_R^z+\sum_{i,k} I_i^{kz}$.
In order to make the numerical effort for a larger number of spins
tractable, we assume a constant HF
coupling for all the nuclei in each dot, allowing us to write the
Overhauser fields as ${\bf h}_i=(\gamma_i /N){\bf I}_i$, $i=L,R$,
where $\gamma_i=\sum_k A_i^k \approx 100$ $\mu$eV is the total HF
coupling, $N$ is the number of nuclei per dot, and ${\bf I}_i$ is
the collective spin operator for dot $i$.  This approximation
provides two more integrals of motion, namely the two $SU(2)$
Casimir operators ${\bf I}_L^2$, ${\bf I}_R^2$, and enables us to
further divide the Hilbert space, making the complexity of the
problem scale polynomially with $N$ instead of exponentially.

We have tested the validity of the uniform HF coupling approximation
for the nuclear polarization dynamics studied here by slicing the
dot into concentric rings. Assigning different HF coupling for each
ring, and assuming no inter-ring dynamics, we find that the dynamics
described below are largely unaffected by this averaging
procedure.\cite{HF1}
Studies of nuclear spin diffusion in a quantum dot also verify that
inter-ring nuclear spin dynamics is significantly suppressed for
smaller quantum dots.\cite{NSD}
Our approach enables us to study the interplay between HF and
exchange effects within a unified theory through exact
diagonalization of the Hamiltonian (\ref{Ham}). The system dynamics
under any applied gate-pulse are calculated without resorting to the
quasi-static approximation that may not be appropriate to describe
nuclear polarization dynamics.\cite{Taylor2} At the same time, our
use of collective spin states enables us to consider a substantially
larger number of nuclei ($\sim 1000$ per dot) as compared with
previous numerical studies, \cite{Shenvi,Schliemann} which is
important in determining the scaling properties of the dynamical
features with $N$.
%

The nuclear spin configuration in each dot is represented in the
basis of collective Dicke states $|j,m\rangle$, where $0\leq j \leq
N/2$ is the total spin of the state (or cooperative number) and $|m|
\leq j$ is the $z$ projection of the total spin. The initial nuclear
spin configuration in the two dots is written as
\begin{equation}
| \psi (0)\rangle_{\rm nuc}\! =\!\sum_{I_L=0}^{N/2} w(I_L)
|I_L,I_L^{z0}\rangle \otimes \! \sum_{I_R=0}^{N/2} w(I_R)
|I_R,I_R^{z0}\rangle, \label{Nuc_state}
\end{equation}
and will be denoted henceforth as $(I_L^{z0},I_R^{z0})_w$ where
$I_i^{z0}$ indicates the initial polarization of the nuclear
configuration in dot $i$. The distribution weights $w(I_i)$ are
assigned statistically by the number of possible ways in which one
can arrange $N$ spins into singlets, triplets, quintets, etc. We
find that these weights obey a Gaussian distribution peaked at
$I_i=\sqrt{N/2}$ whose width is $\sqrt{2N}$, as shown in
Fig.~\ref{Fig1}c.\cite{HF2}
%

An example of the proposed dynamical polarization pulse sequence is
shown in the lower panel of Fig.~\ref{Fig1}a.\cite{Marcus} The key
is to drive the system through the $S-T_+$ resonance with different
speed in the two directions, so that only in one direction can the
electron-nuclear spin flip occur. An $S(0,2)$ state is prepared by
positively detuning the double dot ($P$) to enabe electron exchange
with the leads, whose Fermi level is above the $S(0,2)$ but below
the doubly occupied triplets. The electrons are then separated using
rapid adiabatic passage, where the bias is swept to a negative
detuning quickly relative to the HF coupling but slowly as compared
to the electron tunneling between the hybridized singlet states. Our
simulations show that the adiabaticity requirement is met with sweep
times of 5 ns. The bias is then swept back slowly through the
$S-T_+$ anticrossing ($D$, which can be identified experimentally
\cite{Petta}) where the HF interaction mediates electron spin flip
flop with the nuclear spin baths. Choosing point $S'$ to be far
enough from the $S-T_0$ degeneracy, the electron spin state is
always flipped from a singlet to a $T_+$, thus polarizing the
nuclear spin baths. Finally, the system is swept back to point $P$
where the triplet state relaxes quickly through electron exchange
with the leads, and a new $S(0,2)$ state is prepared for the next
cycle. The singlet preparation at the end of each cycle is simulated
by partially tracing the electronic subsystem and applying a direct
product of the resulting nuclear density matrix with the electronic
configuration at point $P$. To study effects of the polarization
cycles on the relaxation of the electron singlet state, we add a
measurement cycle where the dots are negatively detuned to the
$S-T_0$ degeneracy ($S$) for a separation time $\tau_S$, followed by
a measurement of the singlet probability
$P_S$ ($\tau_M \sim 5 \mu$s).\cite{Johnson,Petta} 
To enhance the efficiency of the polarization process, we perform a
nonlinear bias sweep, spending a substantial part of the cycle in
the vicinity of $D$ (see Fig.~\ref{Fig1}a).

\begin{figure}[htbp]
\epsfxsize=0.85\columnwidth
\vspace*{1.5cm} \centerline{\epsffile{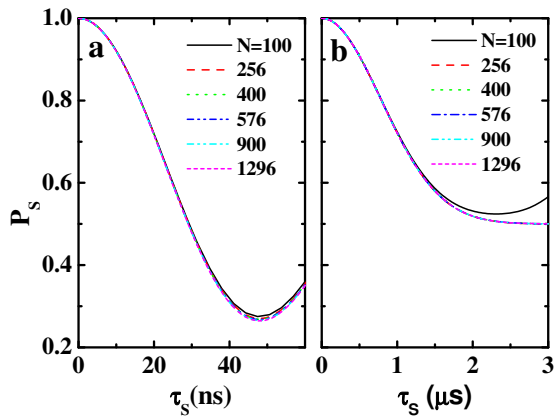}} \vspace{-2.0cm}
\caption{(Color online) Electron singlet probability as a function
of separation time for $N$ values ranging from 100 to 1296, and
$J=0$. (a) Initial nuclear configuration
$(0,-\min[\sqrt{N},I_R])_w$. Time axes are scaled like
$\sqrt{N/10^5}$, so that the presented decay times correspond to
$N=10^5$. The rise in $P_S$ indicates decayed oscillations, which
are an artifact of the uniform coupling approximation and can be
eliminated by performing dynamics averaging over different HF
couplings. (b) Initial nuclear configuration $(-I_L,-I_R)_w$.}
\label{Fig2}
\end{figure}

Before presenting the simulation results for the polarization
scheme, we discuss the envisaged impact of this procedure on the
decay time of the electron singlet correlations.
Figure~\ref{Fig2} depicts the time evolution of $P_S$ when the
electrons are prepared in a singlet state and placed at point $S$,
for several values of $N$.  In Fig.~\ref{Fig2}a the initial nuclear
state is $(0,-\min[\sqrt{N},I_R])_w$, representing a statistical
polarization difference between the two dots. The time axes are
multiplied by $\sqrt{N/10^5}$, indicating that the decay time scales
like $1/\sqrt{N}$.  The decay time of 25 ns agrees well with the
experimental findings,\cite{Petta,Laird} and its scaling with $N$
corresponds to the decoherence time behavior found in
ref.~\onlinecite{Coish}, since we scale the HF coupling constant
with a fixed number of nuclei ($10^5$) rather then with $N$. Similar
dynamics is found for other initial nuclear states, e.g.,
$(-\min[\sqrt{N},I_L],-\min[2\sqrt{N},I_R])_w$. Fig.~\ref{Fig2}b
shows the time evolution of $P_S$ for the fully polarized nuclear
configuration $(-I_L,-I_R)_w$.  In this case $P_S$ decay times are
two orders of magnitude longer reaching $3 \mu$s and they do not
scale with $N$. Other equally polarized configurations do not always
present an enhanced singlet coherence, indicating that the fully
polarized nuclear state is characterized by a narrower distribution
in addition to having $\delta h_z=0$. We stress that the term fully
polarized does not suggest that all the spins are polarized, since
one is limited by the total spin of each collective state within the
weighted distribution. In fact, the total attainable polarization is
$p=1.4533/\sqrt{N}$ reaching a value of 0.46\% for $N=10^5$, which
is consistent with recent experiments.\cite{Marcus,HF3}

The short time dynamics can be understood within a $2 \times 2$
effective Hamiltonian for the $S_z=0$ subspace, $H_{\rm eff} =
\frac{J}{2} (1+\tau_z) +\delta h_z \tau _x$, where $\mbox{\boldmath
$\tau$}$ is the pseudospin operator ($|S\rangle \rightarrow
|\tau_z=-1\rangle$, $|T_0\rangle \rightarrow
|\tau_z=1\rangle$).\cite{Coish} Even when $\langle \delta h_z
\rangle=0$ as for the fully polarized state in Fig.~\ref{Fig2}b, the
spin dynamics do not vanish altogether. The reason is that applying
$H_{\rm eff}$ on each of the collective states in the weighted
distribution results in different eigenvalues and their dynamics do
not cancel out, giving rise to quantum fluctuations. In both cases
shown in Fig.~\ref{Fig2} there is a long time ($\sim 2.5 \mu$s for
$B=100$ mT) envelope decay attributed to higher-order corrections to
$H_{\rm eff}$ that are contained in Eq.~(\ref{Ham}). This envelope
scales like $1/N$ and could therefore govern the dynamics of the
fully polarized state in the large $N$ limit. Its long timescale is
a consequence of the large Zeeman splitting as compared with the
Overhauser fields and it can be made longer using a larger magnetic
field. For the $N$ values we are considering, these corrections are
only observed for the state $(0,0)_w$ for which all other dynamics
are shut down.

Our results also agree well with the experimental findings
\cite{Laird} and analytical results \cite{Coish} for $J>0$. These
include preservation of the singlet correlations over a long
timescale in the limit of $J\gg E_{\rm nuc}$, and the appearance of
damped oscillations in $P_S$ with a saturation value that depends on
$E_{\rm nuc}/J$ in the intermediate regime $J \sim E_{\rm nuc}$. We
also find the long time ($\tau_S \gg T_2^*$) value of $P_S$ to be
$1/2$ for $B \gg B_{\rm nuc}$ and $1/3$ for $B \ll B_{\rm nuc}$, in
agreement with semiclassical theory \cite{Merkulov}.

\begin{figure}[htbp]
\epsfxsize=1\columnwidth
\vspace*{1.9cm} \centerline{\epsffile{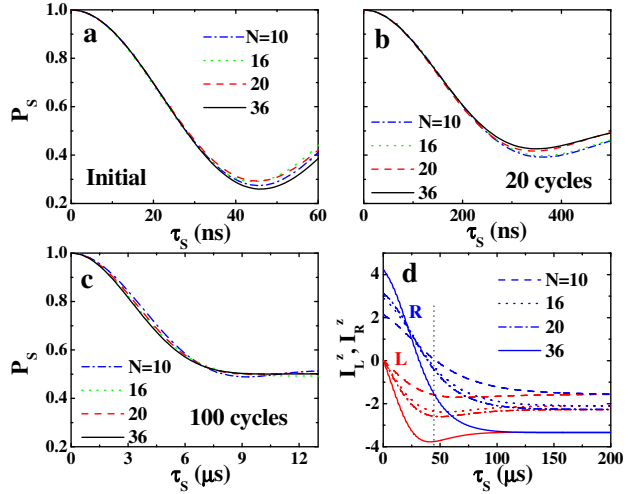}} \vspace{-2.2cm}
\caption{(Color online) Singlet probability vs. separation time,
calculated after performing polarization cycles, for several $N$
values. The initial nuclear configuration is
$(0,-\min[\sqrt{N},I_R])_w$ and $J=0$. (a) Initial $P_S$ without
polarization. Time axes are scaled like $\sqrt{N/10^5}$ (b) $P_S$
calculated after 20 polarization cycles. Time axes are scaled like
$(N/10^5)^{0.28}$ (c) $P_S$ calculated after 100 polarization
cycles. Time axes do not scale with $N$ (d) The corresponding
nuclear polarizations in the left (L-red lines) and right (R-blue
lines) dot. The vertical dotted line corresponds to the time elapsed
after 20 cycles, at which $P_S$ shown in (b) was calculated.}
\label{Fig3}
\end{figure}

Now we investigate the effects of the polarization cycles on the
electron spin states by separating the electrons every four cycles
to calculate $P_S$ as the polarization progresses.  Bias changes
require calculating the evolution separately for each bias, using
the resulting density matrix at each step as the initial condition
for the subsequent step. The numerical effort is thus much more
demanding and we are limited to several tens of nuclei per dot.
Figs.~\ref{Fig3}a-c show $P_S$ calculated for the initial nuclear
configuration, after 20 polarization cycles, and after 100 cycles.
The singlet decay times show a gradual enhancement as the nuclear
polarization builds up, and their scaling with $N$ gradually shifts
from $1/\sqrt{N}$ (Fig.~\ref{Fig3}a) to 1 (Fig.~\ref{Fig3}c).  An
enhancement of factor 300 is obtained for the singlet decay times
that reach $\sim 8 \mu$s when the polarization process is complete.
The corresponding nuclear polarizations in the two dots, shown in
Fig.~\ref{Fig3}d, equilibrate during the polarization process. This
equilibration effect is robust to any of our choices of initial
nuclear configuration, and the degree of equilibration depends on
the symmetry of the double dot. The equilibration between the two
nuclear spin configurations and the narrow distribution of the
Overhauser fields formed during the polarization process
are responsible for the prolonged singlet decay time.

Dipolar interaction between the nuclear spins can break the weighted
distribution of the spin collective states, and compete with the
equilibration process. The number of polarization cycles needed to
complete the process is $n_{\rm cyc}=2\sum_{I=0}^{N/2} w_I^2 I
\approx 1.4533 \sqrt{N}$.  The spin transfer time at point $D$ is
found to be $15.5 \mu {\rm s}/\sqrt{N}$.  As long as the $S-T_+$
degeneracy can be determined accurately, much of the cycle time in
the large $N$ limit is spent to prepare a singlet at $P$. Taking
$t_P=200$ ns for the electron exchange with the leads, the resulting
time to complete the polarization is $\sim 120 \mu$s.

A simple phase space argument seems to suggest that the interdot
Zamboni effect should occur naturally through nuclear spin exchange
between the two baths mediated, e.g., by one or two electrons in the
double dot. While the fully polarized state has a single
configuration, the non-polarized state is highly degenerate having
$C^{N/2}_N$ configurations (assuming equal HF coupling for all the
nuclei). Even for states with statistically similar dot
polarizations the phase space difference could drive an
equilibrating process. For example, the ratio between the number of
configurations in $(0,\sqrt{N})_w$ and $(\sqrt{N}/2,\sqrt{N}/2)_w$
is $e^{-1/4}$ which, as our simulations show, is sufficient to
induce an equilibrating process.
We have performed simulations for the singlet decay
times after a nuclear preparation time $\tau_{\rm eq}$ (on the order
of tens of $\mu$s) in which we introduce one or two electrons into
the double dot with appropriate bias and magnetic field.
Interestingly, in spite of the equilibration of the two nuclear
configurations, the singlet decay times are only modestly extended.
This is because the distribution over $I_i^z$ of the collective
nuclear states is broadened during these naturally occurring
electron-mediated equilibration processes. The time scale for this
nuclear dynamics is in the order of $h N E_Z/\gamma_i^2 \approx 20$
ns, comparable to that for the electron spin dynamics, so that
quasi-static approximation for the nuclear quantities becomes
invalid in the regime of $\langle \delta h_z \rangle \approx 0$. The
resulting fluctuations suppress the singlet decay time enhancement
even though $\delta h_z$ becomes small. In contrast, such dynamics
does not exist in the fully polarized collective states, where the
pumping of singlet states pushes $I_i^z$ to their minimum values.
The polarization cycles, while unable to produce high degrees of
polarization, significantly regulate the nuclear spin states and
reduce the nuclear field fluctuations, thus extend the decay time of
the singlet state.

In summary, we have studied a nuclear polarization scheme in gated
double dots utilizing the $S-T_+$ degeneracy point, and examined its
impact on the two-electron spin singlet decay time, obtaining
two-orders-of-magnitude enhancement.  We have shown that high degree
of nuclear polarization is not essential to suppress the nuclear
relaxation channel for a two-electron spin state.  Instead,
enhancement of the singlet relaxation time is obtained by
electron-mediated equilibration process within the two nuclear baths
that suppresses the Overhauser field fluctuations within each
reservoir. We have explored other strategies to facilitate this
equilibrating process, and have found that while equal nuclear
polarizations between the two dots may be obtained, they are not
accompanied by narrowing of the nuclear state distribution and thus
do not result in a dramatic enhancement in the singlet decay time.

We thank C.~M.~Marcus and J.~R.~Petta for very useful discussions.
This work is supported by NSA/LPS and ARO.


\begin{thebibliography}{99}
%
\bibitem{ZFD} I.~Zutic, J.~Fabian, and S.~Das Sarma, Rev.~Mod.~Phys. {\bf 76},
323 (2004).
%
\bibitem{Johnson} A.~C.~Johnson {\it et al.},
Nature {\bf 435}, 925 (2005); R.~Hanson {\it et al.},
Phys.~Rev.~Lett.~{\bf 94}, 196802 (2005).
%
\bibitem{Koppens} F.~H.~L.~Koppens {\it et al.},
Science {\bf 309}, 1346 (2005).
%
\bibitem{Petta} J.~R.~Petta {\it et al.},
Science {\bf 309}, 2180 (2005).
%
\bibitem{Merkulov} I.~A.~Merkulov, Al.~L.~Efros, and M.~Rosen
Phys.~Rev.~B {\bf 65}, 205309 (2002).
%
\bibitem{Schliemann} J.~Schliemann, A.~V.~Khaetskii, and D.~Loss,
Phys.~Rev.~B {\bf 66}, 245303 (2002).
%
%
\bibitem{Shenvi} N.~Shenvi, R.~de Sousa, and K.~B.~Whaley,
Phys.~Rev.~B {\bf 71}, 224411 (2005).
%
\bibitem{Coish} W.~A.~Coish and Daniel Loss, Phys.~Rev.~B {\bf 72},
125337 (2005).
%
\bibitem{Deng} C.~Deng and X.~Hu, Phys.~Rev.~B {\bf 73},
241303(R) (2006).
%
\bibitem{Yao} W.~Yao, R.~B.~Liu, and L.~J.~Sham,
Phys.~Rev.~B {\bf 74}, 195301 (2006).
%
\bibitem{Klauser} D.~Klauser, W.~A.~Coish, and Daniel Loss, Phys.~Rev.~B {\bf
73}, 205302 (2006); D.~Stepanenko, G.~Burkard, G.~Giedke, and
A.~Imamoglu, Phys.~Rev.~Lett.~{\bf 96}, 136401 (2006).
%
%
\bibitem{Burkard} G.~Burkard, D.~Loss, and D.~P.~DiVincenzo,
Phys.~Rev.~B {\bf 59}, 2070 (1999).
%
\bibitem{Taylor1} J.~M.~Taylor, C.~M.~Marcus, and M.~D.~Lukin,
Phys.~Rev.~Lett.~{\bf 90}, 206803 (2003).
%
\bibitem{Gammon} A.~S.~Bracker {\it et al.},
Phus.~Rev.~Lett.~{\bf 94}, 047402 (2005).
%
\bibitem{Ono} K.~Ono and S.~Tarucha, Phys.~Rev.~Lett.~{\bf
92}, 256803 (2004).
%
\bibitem{Marcus} J.~R.~Petta and C.~M.~Marcus (unpublished).
%
\bibitem{Sousa} R.~de Sousa and S.~Das Sarma, Phys.~Rev.~B {\bf 68}, 115322
(2003).
%
%
\bibitem{Zamboni} Zamboni Machine, named after its inventor Frank J.~Zamboni,
has become a generic name for resurfacing ice rink machines. The
analogy to the supression of quantum fluctuations in the nuclear
reservoirs was suggested by C.~M.~Marcus.
%
\bibitem{HD} X.~Hu and S.~Das Sarma, Phys.~Rev.~A {\bf 61}, 062301 (2000).
%
%
\bibitem{HF1} Inter-ring dynamics would result in the center rings being
polarized first. The outer rings, for which the nuclear dynamics is
slower, will tend to depolarize the nuclei in the center.
Nevertheless, once the polarization is complete, all of the rings'
collective nuclear states will be polarized, resulting in a
qualitatively similar distribution to the one that is considered in
the main text. Further study is needed to quantify the effects of
inter-ring dynamics.
%
\bibitem{NSD} C.~Deng and X.~Hu, Phys.~Rev.~B {\bf 72}, 165333 (2005).
%
\bibitem{Taylor2} J.~M.~Taylor, J.~R.~Petta, A.~C.~Johnson, A.~Yacoby,
C.~M.~Marcus, and M.~D.~Lukin, cond-mat/0602470 (unpublished).
%
\bibitem{HF2} This nuclear state distribution is in fact wider than the
distribution obtained when considering non-uniform HF couplings, the
latter given by a sum of Gaussian distributions corresponding to the
number of nuclei in each ring.
%
\bibitem{Laird} E.~A.~Laird {\it et al.},
Phys.~Rev.~Lett.~{\bf 97} 056801 (2006).
%
\bibitem{HF3} Slicing the dot to 100 rings gives $p=4.3\%$ for $N=10^5$.
The HF coupling differences within each ring are then smaller than
0.1\%, roughly preserving the total spin numbers in each ring.
%
%
%
%
\end{thebibliography}
\end{document}